\documentstyle[aps]{revtex}
\begin{document}
\draft
\title{Observational Evidence Against Birefringence over Cosmological 
Distances}

\author{J.F.C. Wardle}
\address{Physics Department, Brandeis University, Waltham MA 02254}

\author{R.A. Perley}
\address{National Radio Astronomy Observatory, P.O. Box 'O',
Socorro, NM, 87801} 

\author{M.H. Cohen} 
\address{Astronomy Department,
California Institute of Technology, Pasadena, CA 91125}
 
\date{\today}

\maketitle

\begin{abstract}

We show that recent radio and optical observations of polarized
radiation from well-resolved high redshift quasars and radio galaxies
rule out the cosmological rotation of the plane of polarization
claimed recently by Nodland and Ralston.  A least squares fit to the
radio data has a slope only 2\% of their claimed effect.

\end{abstract}


In a recent paper, Nodland and Ralston \cite{n97} claim to find a
systematic rotation of the plane of polarization of electromagnetic
waves propagating over cosmological distances. Their claimed effect is
large, requiring the plane of polarization from high redshift objects
to be rotated by as much as 3.0 radians -- an easily detectable
signature.  Here we report new optical data taken with the Keck
Telescope, and radio observations made with the Very Large Array (VLA)
which show that any such rotation is less than 3 degrees out to
redshifts in excess of two.

The data used by Nodland and Ralston consisted of radio measurements
of the {\sl integrated} polarization of extragalactic radio
sources. After correcting for Faraday rotation, they compare the
intrinsic angle of polarization, $\chi$, to the major axis of the
radio source, $\psi$. They suggest that ``On symmetry grounds $\chi$
would be expected to align with the major axis angle $\psi$ of the
galaxy.''  High resolution observations of extragalactic radio sources
\cite{b92,b94} show that such an expectation is optimistic. The
integrated polarization of a radio source is the vector sum of the
polarized radiation from several different emission regions, which
have different angles of polarization.  The resulting degree of
polarization is generally low, with a net angle only weakly related to
the major axis of the radio source.

Nodland and Ralston fit the misalignment angle $\beta = \chi - \psi$
to a dipole anisotropy of the form $\beta = \frac{1}{2}
\Lambda_{s}^{-1} r \cos (\gamma)$. Here, $r$ is a distance which they
define as $r = 6.17 \times 10^{25}[1-(1+z)^{-\frac{3}{2}}]$ meters for
a Hubble constant of 100 km/sec/Mpc, and $\Lambda_s$ is a scale length
which they find to be about $0.9 \times 10^{25}$ meters, or nearly one
billion light years. $\gamma$ is the angle between the direction to
the radio source and a pole direction $\vec{s}$, whose coordinates are
$20 \pm 2$ hours in right ascension, and $-10 \pm 20$ degrees in
declination.  Thus, the Nodland and Ralston relation states that the
plane of polarization of emission from cosmological objects should be
seen to be rotated by an amount: $\beta = 31.8 r \cos(\gamma)$
degrees.  Note that in their paper, Nodland and Ralston arbitrarily
add $\pm \pi$ to the measured $\beta$ values such that $\beta$ is
constrained to be positive for positive $\cos\gamma$, and negative for
negative $\cos\gamma$.  It is this procedure which leads at first
sight to the apparently strong correlation in their Fig. 1d.  In fact,
judgements of the significance must be based on the data within each
quadrant separately, as is pointed out by NR.   

We shall not discuss the statistical or theoretical arguments in their
paper.  Problems with their statistical methods have been discussed by
Carroll and Field \cite{cf97}, Eisenstein and Bunn \cite{e97}, and
Loredo, Flanagan and Wasserman \cite{lf97}.  Here we will appeal
directly to recent high resolution polarization observations of
distant objects with well defined structural axes against which we can
test the claimed rotation. A cosmological rotation of this type has
already been rejected by Cimatti {\em et al.} \cite{c94} based on
polarized UV light from the distant radio galaxy MRC 2025-218 ($z =
2.63$). The polarized light is due to electron or dust scattering of
light emanating from the galaxy nucleus, and the observed electric
vectors are nearly exactly perpendicular to the axis of extended UV
and Ly$\alpha$ emission, as is often found in high redshift radio
galaxies \cite{c94,c93,d96}. Cimatti {\em et al.} \cite{c94} show that the
angle $\chi - \psi = 87 \pm 10^{\circ}$, and state that ``the plane of
polarization is not rotated by more than ten degrees when the
radiation travels from $z = 2.63$ to us.'' The rotation predicted from
the relation of Nodland and Ralston for this object is $163^{\circ}$
-- close to a $\pi$ ambiguity, however. The presence of such an
ambiguity can be confidently ruled out by the results of optical
polarimetry for lower redshift radio galaxies,\cite{c93,d96}, which
show a strong tendency for the misalignment angles to cluster near
$90^{\circ}$.

With a large optical telescope, a polarization image of a faint,
distant radio galaxy can be made, and this directly gives the
polarization distribution of the radiation from the extended emission
regions. When this is done, the polarization vectors typically show a
centro-symmetric pattern, with the E-vectors perpendicular to the
radius to the nucleus. This is the signature of scattering (whether by
dust or from electrons) of radiation from a point source.  A
circularly symmetric pattern in such an image is clear evidence of an
absence of rotation of the plane of polarization (provided there is
not an $n\pi$ rotation of the plane of polarization to that particular
object).

Recent optical observations with the Keck telescope by Cohen and
collaborators \cite{c97} have produced polarization images with low
noise and tight upper limits on any rotation. In Fig.\ \ref{OptExamp}
we show the results for 3C\,265 -- a radio galaxy with a redshift of
$z = 0.811$ for which the expected rotation from Nodland and Ralston
is $-57^{\circ}$.  Our analysis of the data gives a $3-\sigma$ upper
limit for any anomalous rotation of $<4^{\circ}$. The same result
comes from observations of 3C\,405 (= Cygnus A), where the `expected'
rotation is $11^{\circ}$, but the limit is observed to be $5^{\circ}$.
Several other radio galaxies with redshifts near 0.1 also show no
rotation to a limit of about $3^{\circ}$ (Cohen, priv. comm.)

An equally stringent test of cosmological birefringence at radio
wavelengths comes from VLA observations of the jets in powerful radio
quasars.  It is well established \cite{bp84,b84} that in high luminosity
radio jets, the magnetic field is closely aligned with the local jet
direction, so that the plane of polarization of the radio waves
(emitted by the synchrotron process) is perpendicular to the jet. Such
jets therefore provide well defined reference directions, $\psi$,
against which we can measure any cosmological rotation in the plane of
polarization.  Note that this is not the only well-defined structural
axis with which an accurate local measure of the misalignment angle
can be made.  Leahy \cite{l97} uses the well-known alignment of
polarization with the sharp intensity boundaries seen in the lobes of
radio galaxies and quasars in his response to \cite{n97}.

In Fig.\ \ref{Radio} we show an image of the radio quasar PKS 2209+152
($z = 1.502$) made with the VLA at $\lambda$ 3.6 cm. The contours are
total intensity, showing the compact central core of the quasar to the
south-west (lower right) and a thin jet curving to the north-east,
ending in a bright compact ``hotspot''.  The degree of polarization is
between 20\% and 50\% throughout the jet and hotspot.  The tick marks
show the plane of polarization of the radio waves, $\chi$, and it is
seen that the polarization vectors are closely orthogonal to the jet
direction over most of its path. (It is also easily seen that the {\em
integrated} polarization of this quasar will bear no particular
relationship to the major axis of the source.)  The relation published
by Nodland and Ralston predicts a rotation of the plane of
polarization of about $115^\circ$ -- but it is clear from this figure
that such a rotation is not present.  We emphasize that the relation
seen in this figure is faithfully reproduced in all radio observations
of radio quasars known to us that have been made to date.
 
To demonstrate quantitatively the absence of any cosmological rotation
of the plane of polarization, we use two samples of quasars for which
deep VLA images at high resolution are available. Both samples were
selected on the basis of the strength and large angular extent of
their radio structure -- ideal criteria for testing for any
cosmological rotation of the plane of polarization.  The first is from
Bridle {\em et al.} \cite{b94}, and contains thirteen bright quasars
of large angular size from the 3CR Catalog \cite{l83}, observed at
$\lambda$ 6 cm. We omit from our analysis 3C\,215 and 3C\,249.1, whose
jets are short and distorted, and 3C\,432, whose jet is
unpolarized. The second sample consists of ongoing observations of
nineteen high redshift quasars with bright jets made by Kronberg,
Perley, R\"oser and Dyer (unpublished) at $\lambda$ 3.6 cm. The
$\lambda$ 6 cm data were corrected for Faraday rotation, using
published rotation measures \cite{k81}. The $\lambda$ 3.6 cm data were
not corrected, as only twelve of the objects have a published rotation
measure \cite{k81,g91}.  However, for these twelve objects, the mean
correction at this wavelength is less than $4^\circ$, and there is no
indication that the corrections for the others will be any larger.
These corrections are far smaller than the effect we are searching
for.  For each quasar, we measure the misalignment angle, defined as
$\chi - \psi$, using only the integrated polarization of a straight,
well-resolved portion of each jet.  Following Carroll and Field, we
constrain this difference to lie between 0 and 180 degrees.  We note
that seven of the ten Bridle {\sl et al.} quasars, and three of the
nineteen Kronberg {\sl et al.} quasars are in the sample used by
\cite{n97}, and that all of our sample have $z > 0.3$.

The results for the 29 quasars are shown in Fig.\ \ref{Data}, where we
plot $\chi - \psi$ against $r\cos\gamma$. We also show the relation
proposed by Nodland and Ralston. It is obvious that the radio and
optical data directly refute such a relation. A regression line fitted
to our data has a slope of $0.54 \pm 0.74$ degrees per $10^{25}$ m,
less than 2\% of Nodland and Ralston's slope of about 32 degrees per
$10^{25}$ m. Our $3\sigma$ upper limit is fourteen times smaller than
their slope. The scatter in the radio data can be attributed entirely
to noise and to small deviations in the local path of the jets.  

The zero intercept for our fit is $89.6^{\circ} \pm 1.7^{\circ}$, as
expected for synchrotron emission where the magnetic field is oriented
along the emitting structure -- the jet.  Note that this offset from
the relation proposed by Nodland and Ralston is arbitrary -- we could
have equally well defined our misalignment angle with respect to the
normal to the observed E-vector, in which case the intercept would
have been $-0.4^{\circ}$.  The physical significance lies in the slope
of the relation.

In summary, the observational data taken with modern high resolution
instruments of high redshift galaxies and quasars at both optical and
radio wavelengths show that any rotation of the plane of polarization
(circular birefringence) over cosmological distances is at least a
factor of 50 smaller than that claimed by Nodland and Ralston, and is
statistically indistinguishable from zero.

\bigskip

The National Radio Astronomy Observatory is a facility of the National
Science Foundation operated under cooperative agreement by Associated
Universities, Inc. The W.M. Keck Observatory is a scientific
partnership between the University of California and the California
Institute of Technology, made possible by the generous gift of the
W.M. Keck Foundation and support of its president, Howard Keck.  We
thank P. P. Kronberg, D. S. De Young and B. F. Burke for stimulating
discussions, and P. P. Kronberg, H.-J. R\"oser, and C. C. Dyer for
permitting the use of their data in advance of publication.  JFCW
acknowledges support from NSF grant AST9224848.  We are grateful to
Borge Nodland for assistance in identifying sources common to our
samples, and for general discussions.

\begin{figure}
\caption{Optical $V$ band imaging polarimetry of 3C265, a radio
galaxy with $z = 0.811$ for which the relation of Nodland and Ralston
predicts a rotation of $-66^{\circ}$. Contours are plotted at 4, 8,
16, 32 and 64\% of the peak brightness.  The position angle and degree
of polarization of the polarized light are given by the orientation
and length of the plotted vectors, with the rightmost vector
representing 22.3\%.  Only vectors with an error less than $7^{\circ}$
are shown.  The centrosymmetric pattern of vectors is caused by
scattering of light from a hidden central source. The mean deviation
of the 53 plotted vectors from the perpendicular to a line joining
each to the nucleus is $-1.4^{\circ} \pm 1.1^{\circ}$.}
\label{OptExamp}
\end{figure}

\begin{figure}
\caption{A map made with the VLA at a wavelength of 3.6 cm of the jet in the
radio-loud quasar PKS 2209+152, ($z = 1.502$) with a resolution of
$0.25''$.  The relation of Nodland and Ralston predicts a rotation of
the plane of polarization of $113^{\circ}$.  The curved structure, and
high degree of polarization is typical for objects of this type.
There is a negligible rotation of the position angle of the polarized
radiation from that expected on the basis of observations of nearby
objects of the same class.}
\label{Radio}
\end{figure}

\begin{figure}
\caption{A plot of the measured deviation between the local jet
direction and the position angle of polarized emission for 29 high
redshift quasars from two different surveys.  Superposed is the
relation claimed by Nodland and Ralston.  The detailed data do not
support their claim of a cosmological rotation in the plane of
polarization of high redshift objects.}
\label{Data}
\end{figure}

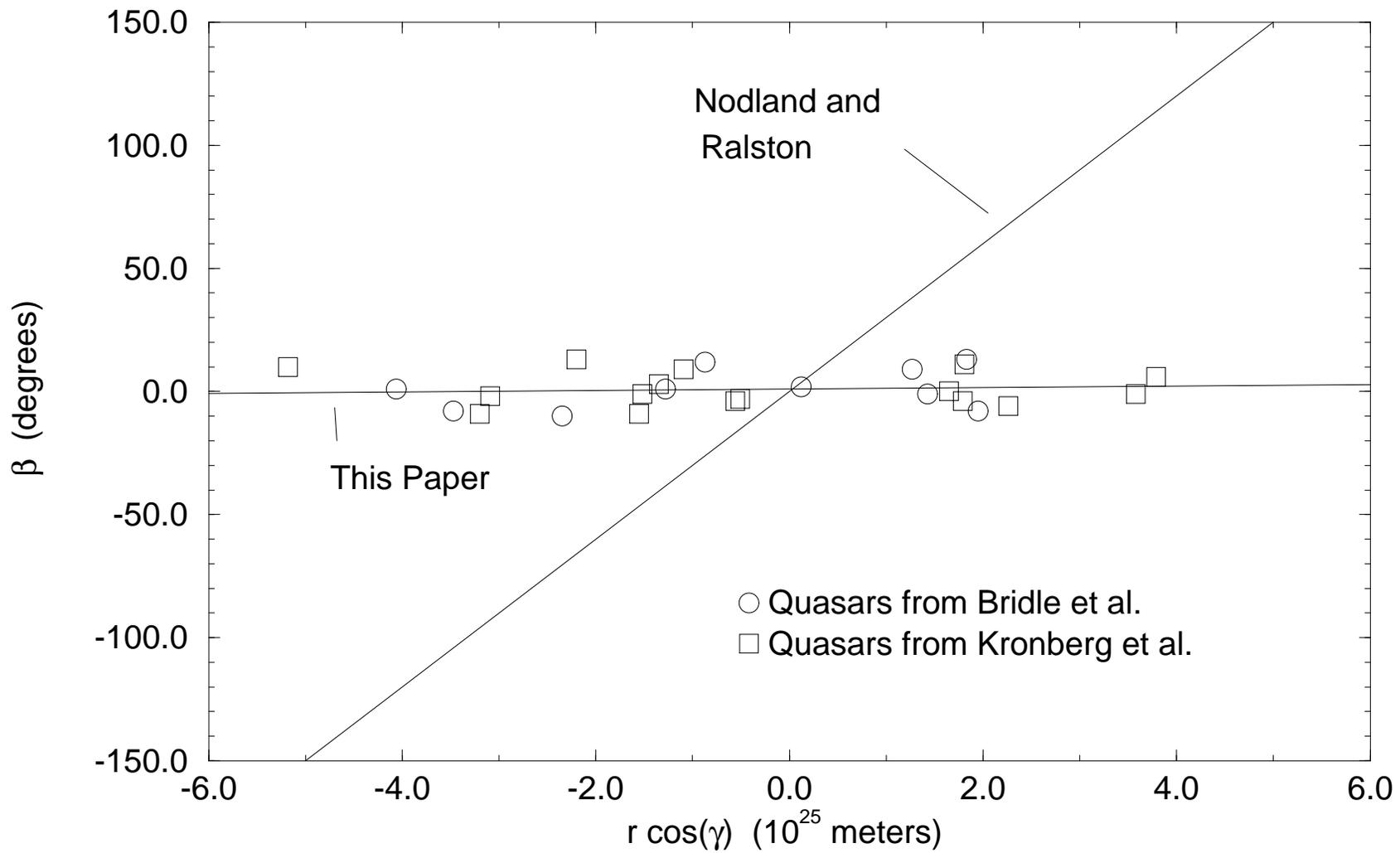
\end{document}